\documentclass[conference]{IEEEtran}
\IEEEoverridecommandlockouts

\usepackage{cite}
\usepackage{amsmath,amssymb,amsfonts}
\usepackage{amsthm}
\usepackage{algorithmic}
\usepackage{graphicx}
\usepackage{textcomp}
\usepackage{xcolor}

\usepackage{stfloats} 
\usepackage{subcaption}

\usepackage{bbm}

\usepackage{algorithm}
\usepackage{array}
\usepackage[caption=false,font=normalsize,labelfont=rm,textfont=rm]{subfig}

\def\BibTeX{{\rm B\kern-.05em{\sc i\kern-.025em b}\kern-.08em
    T\kern-.1667em\lower.7ex\hbox{E}\kern-.125emX}}
\usepackage{wrapfig}
\usepackage{afterpage}

\usepackage{subcaption}
\begin{document}

\title{Conformal Robust Beamforming via Generative Channel Models\vspace{-10pt}\\
}

\author{
	\IEEEauthorblockN{Xin~Su\textsuperscript{*}, Qiushuo~Hou\textsuperscript{\dag}, Ruisi~He\textsuperscript{*}, and Osvaldo~Simeone\textsuperscript{\ddag}}
	\IEEEauthorblockA{\textsuperscript{*}School of Electronics and Information Engineering, Beijing Jiaotong University, Beijing, China}
	\IEEEauthorblockA{\textsuperscript{\dag}College of Information Science and Electronic Engineering, Zhejiang University, Hangzhou, China}
	\IEEEauthorblockA{\textsuperscript{\ddag}KCLIP Lab, CIIPS, Department of Engineering, King's College London, London, UK}
	\IEEEauthorblockA{E-mail: xin$\_$su@bjtu.edu.cn}
	\vspace{-32pt} 
\thanks{This work was supported by the National Natural Science Foundation of China under Grant 62431003. The work of O. Simeone and Q. Hou was supported in part by the European Union’s Horizon Europe project CENTRIC (101096379). The work of O. Simeone was also supported by the Open Fellowships of the EPSRC (EP/W024101/1) and by the EPSRC project (EP/X011852/1).}
}


\maketitle

\begin{abstract}
Traditional approaches to outage-constrained beamforming optimization rely on statistical assumptions about channel distributions and estimation errors. However, the resulting outage probability guarantees are only valid when these assumptions accurately reflect reality. This paper tackles the fundamental challenge of providing outage probability guarantees that remain robust regardless of specific channel or estimation error models. To achieve this, we propose a two-stage framework: \textit{(i)} construction of a channel uncertainty set using a generative channel model combined with conformal prediction, and \textit{(ii)} robust beamforming via the solution of a min-max optimization problem. The proposed method separates the modeling and optimization tasks, enabling principled uncertainty quantification and robust decision-making. Simulation results confirm the effectiveness and reliability of the framework in achieving model-agnostic outage guarantees.
\end{abstract}

\vspace{-5pt}
\section{Introduction}
\afterpage{%
	\begin{figure}[H]
		\centering
		\includegraphics[width=0.65\columnwidth]{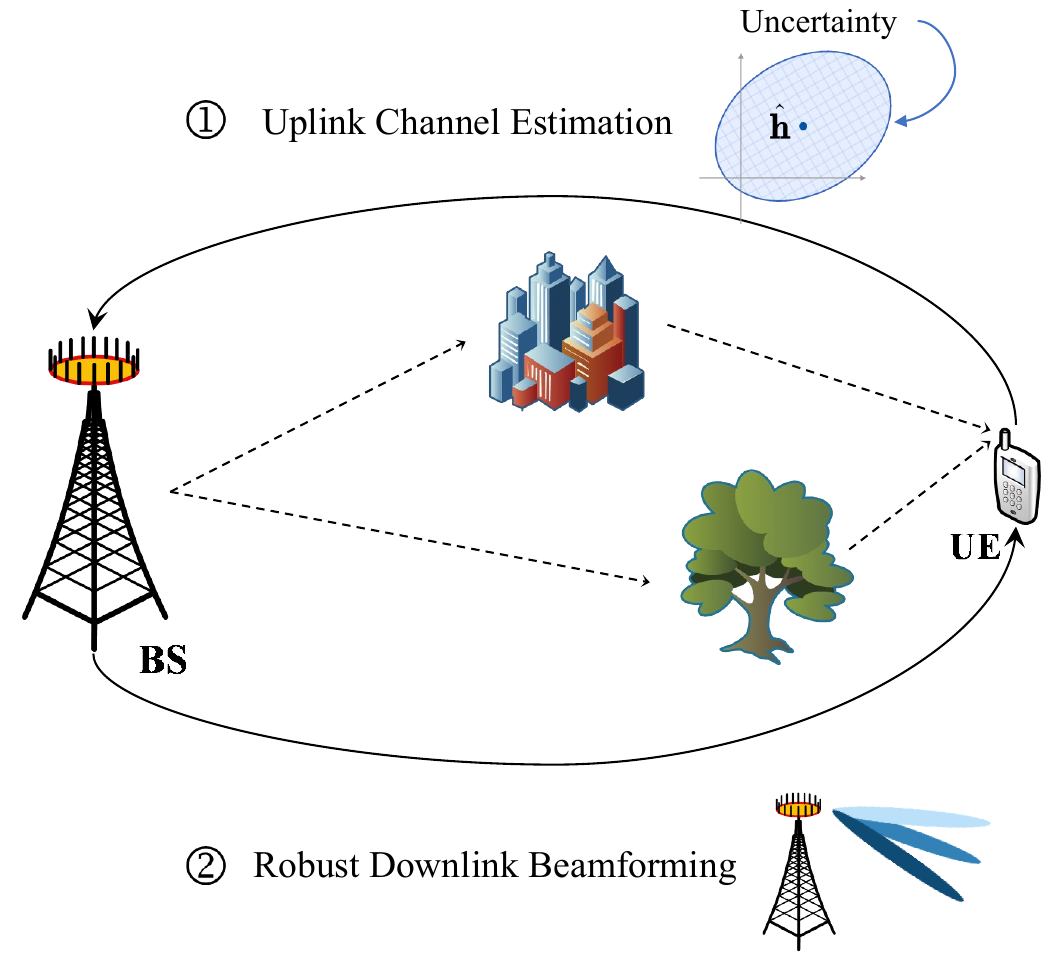}
		\caption{Illustration of a TDD MISO communication system. This paper focuses on the joint optimization of transmission rate and beamforming vector with the goal of ensuring robustness to the inherent uncertainty caused by uplink channel estimation errors.}
		\label{fig:system_model}
		\vspace{-7pt}
	\end{figure} 
}
	
\IEEEPARstart{T}{he} emerging applications such as the industrial Internet of Things and autonomous vehicles, impose diverse and stringent quality-of-service requirements \cite{pourkabirian2021robust}. Meeting these requirements necessitates the design of effective transmission schemes that ensure sufficiently low outage probabilities, in line with the performance demands of these systems. A key challenge in achieving this goal stems from the inevitable uncertainty in the available channel state information (CSI) due to estimation errors. 

A fundamental problem in this context is that of jointly optimizing transmission rate and beamforming vector based on imperfect CSI, while ensuring compliance with outage probability constraints. This robust beamforming optimization problem has been extensively studied over the past decades \cite{chalise2007robust}. As exemplified by  \cite{xiang2012robust,lin2020deep}, a common approach is to assume that the CSI error lies within a convex bounded region. Under this assumption, the problem becomes a convex optimization task that can be efficiently solved using standard methods.

While the bounded uncertainty model is mathematically convenient, it fails to capture the inherently probabilistic nature of wireless fading and CSI estimation errors. To overcome this limitation, statistical models---typically Gaussian---have been widely adopted to model CSI uncertainty.
For example, reference \cite{park2012outage} assumes circularly symmetric complex Gaussian errors and derives closed-form outage expressions. To reduce computational overhead, the work \cite{liang2024data} proposes a data-driven method that transforms the probabilistic outage constraint into a quantile estimation problem. Gaussian error models have also been utilized in other wireless scenarios, including intelligent reflecting surface-assisted networks \cite{zhao2021outage} and integrated sensing and communication systems \cite{liu2022outage}.

In contrast to this body of work, the present paper addresses the fundamental challenge of ensuring outage probability guarantees that remain valid without relying on any specific assumption about the distribution of channel errors. Our approach builds on recent advances in optimization with assumption-free probabilistic guarantees, originally developed in the context of robotics and control systems \cite{zecchin2024forking,lindemann2023safe}. These methods are grounded in conformal prediction (CP), a powerful technique that transforms the output of any black-box predictor into a prediction set with guaranteed coverage properties \cite{angelopoulos2021gentle}. CP has been recently applied to wireless systems for resource allocation \cite{cohen2023guaranteed}, prediction \cite{cohen2023calibrating, zecchin2025generalization}, and beamforming selection \cite{deng2025scan}.

As illustrated in Fig. \ref{fig:system_model}, the proposed framework consists of two steps. First, we construct a reliable channel uncertainty set using a generative channel model \cite{baur2024leveraging} in conjunction with CP. This results in a prediction set that provably contains the true channel with a user-defined confidence level, independently of the underlying error distribution. In the second step, we solve a min-max robust beamforming problem over this data-driven uncertainty set \cite{xiang2012robust}. Simulation results demonstrate that the proposed framework reliably satisfies outage constraints and yields substantial gains in average system throughput.

The remainder of the paper is structured as follows. Section \ref{sec:problem_formulation} formulates the outage-constrained robust beamforming problem. Section \ref{sec:conformal_robustBF} introduces the conformal robust beamforming framework. Sections \ref{sec:CE_CP} and \ref{sec:robustBF} detail the proposed two-step approach. Section \ref{sec:numerical_results} presents numerical results, and Section \ref{sec:conclusion} concludes the paper.

\section{Problem Formulation} 
\label{sec:problem_formulation}
\subsection{Signal Model}
As illustrated in Fig. \ref{fig:system_model}, we consider the problem of robust downlink beamforming design for a time-division duplexing (TDD)  multiple-input single-output (MISO) communication system. A single-antenna user equipment (UE) first transmits pilots to an $N$-antenna base station (BS) for uplink channel estimation. 
Using the channel reciprocity in TDD systems, the BS then utilizes the estimated CSI to optimize the beamforming vectors for downlink transmission.

\subsubsection{Channel Vector}
Let $\mathbf{h}\in \mathbb{C}^{N \times 1}$ denote the spatial channel between BS and UE for a given subcarrier. In this work, we do not make any assumptions about the distribution of channel $\mathbf{h}$, except that it varies in an independent and identically distributed (i.i.d.) manner over some period of time.

\subsubsection{Uplink Channel Estimation}
To estimate the channel $\mathbf{h}$, the UE transmits a known training symbol $s\in \mathbb{C}$ to the BS with normalized power constraint $\mathbb{E} [ \left| s \right|^2] =1$. The noisy signal $\mathbf{y} \in \mathbb{C}^{N \times 1}$ received by the BS is modeled as
\begingroup
\setlength{\abovedisplayskip}{3pt}
\setlength{\belowdisplayskip}{3pt}
	\begin{equation}
		\label{equ:simo_ce}
		\mathbf{y} = \mathbf{h}s + \mathbf{n},
	\end{equation}
\endgroup
where $\mathbf{n} \sim\mathcal{CN}(\mathbf{0},\gamma^2\mathbf{I})$ is complex additive white Gaussian noise (AWGN). Multiple training symbols may also be transmitted to improve the signal-to-noise ratio (SNR) in \eqref{equ:simo_ce}.

\subsubsection{Downlink Beamforming}
Given the received pilot signals in \eqref{equ:simo_ce}, the BS designs a beamforming vector $\mathbf{w}\in \mathbb{C}^{N \times 1}$ to transmit a data symbol $x\in \mathbb{C}$ with normalized power constraint $\mathbb{E}[ \left| x \right|^2] =1$ to the UE. Accordingly, the signal $z \in \mathbb{C}$ received by the UE is given by
\begin{equation}
	\label{equ:simo_bf}
	z=\mathbf{h}^{\mathsf{H}}\mathbf{w}x+n,
\end{equation}
with AWGN $n \sim \mathcal{CN}(0,\sigma^2)$. The power constraint during data transmission is accounted for by the squared norm of the beamforming vector $\left\| \mathbf{w} \right\| ^2$ via the inequality $\left\| \mathbf{w} \right\| ^2\le P$.

\subsection{Problem Formulation}
The design objective is to optimize the beamforming vector $\mathbf{w}$ so as to maximize the system achievable rate, while ensuring that the outage probability remains below a predefined level $\alpha$. Mathematically, the problem is formulated as 

\begingroup
\setlength{\abovedisplayskip}{2pt}
\setlength{\belowdisplayskip}{2pt}
\begin{subequations}
\label{equ:problem_init}
\begin{align}
\label{equ:obj_init}
\underset{\mathbf{w},\bar{R}>0}{\max} \; &\, \bar{R} \\
\label{equ:outage_constraint}
\mathrm{s}.\mathrm{t}. \; &\, {\mathrm{Pr}\left[ R\left( \mathbf{w},\mathbf{h} \right) <\bar{R} \right] }<\alpha  \\
\label{equ:power_constraint}
&\left\| \mathbf{w} \right\| ^2\le P,
\end{align}	
\end{subequations}
\endgroup
where 
\begin{equation}
	\label{equ:rate}
	R\left( \mathbf{w},{\mathbf{h}} \right) =\!\log _2\left( \!1+\frac{\left| {\mathbf{h}}^{\mathsf{H}}\mathbf{w} \right|^2}{\sigma ^2}\! \right)
\end{equation}	
is the rate achieved using beamforming vector $\mathbf{w}$ on channel $\mathbf{h}$. The probability in \eqref{equ:outage_constraint} is evaluated with respect to the true channel $\mathbf{h}$ and to any information used in the evaluation of the beamforming vector $\mathbf{w}$, including the pilot signals in \eqref{equ:simo_ce}.

Prior works have addressed problem \eqref{equ:problem_init} under statistical assumption on the channel $\mathbf{h}$ and on the estimation error. The channel is often modeled following a correlated Rayleigh distribution, while the channel estimation error is assumed to be additive and Gaussian.
Unlike these prior works, we aim to solve problem \eqref{equ:problem_init} without assuming any statistical model for channel and estimation error. 

\vspace{-5pt}
\section{Conformal Robust Beamforming}
\label{sec:conformal_robustBF}
The proposed approach is based on a decomposition of problem \eqref{equ:problem_init} into two steps, namely: (i) channel estimation uncertainty set evaluation, and (ii) min-max robust beamforming. This two-step approach was studied in \cite{zecchin2024forking,lindemann2023safe} for control problems and robotics, and it has been recently shown to be optimal for assumption-free probabilistic constraints such as \eqref{equ:outage_constraint}\cite{lindemann2023safe}.
\subsection{Channel Estimation Uncertainty Set Evaluation}
The first step is to use the pilot information $\mathbf{y}$ in \eqref{equ:simo_ce} to produce an uncertainty set $\mathcal{H} \subseteq \mathbb{C}^N$ in the space of channel vectors with the following coverage property
\begin{equation}
\label{equ:not_in_set}
\mathrm{Pr}\!\left[ \mathbf{h}\notin \mathcal{H} \right] <\alpha ,
\end{equation}
where $\alpha$ is the target probability in the constraint \eqref{equ:outage_constraint}. When the condition \eqref{equ:not_in_set} is satisfied, the set $\mathcal{H}$ contains the true channel $\mathbf{h}$ with probability no lower than $1-\alpha$.

A set $\mathcal{H}$ satisfying the property \eqref{equ:not_in_set} is obtained via an application of CP as detailed in Section \ref{sec:CE_CP}.

\subsection{Min-Max Robust Beamforming}
Using the uncertainty set $\mathcal{H}$, the beamforming vector is obtained by addressing the min-max problem
\begin{subequations}
\label{equ:problem_robustbf}
\begin{align}
\label{equ:obj_robustbf}
\max_{\mathbf{w}} \min_{{\mathbf{h}}\in \mathcal{H}} &\left|{\mathbf{h}}^{\mathsf{H}}\mathbf{w} \right|^2
\\
\label{equ:constraint_power}
\mathrm{s}.\mathrm{t}.\,&\left\| \mathbf{w} \right\| ^2\leq P.
\end{align}
\end{subequations}

Denote as $\left( \mathbf{w}^{\star},\mathbf{h}^{\star} \right) $ a solution to problem \eqref{equ:problem_robustbf}. Thanks to the coverage property \eqref{equ:not_in_set} of the uncertainty set $\mathcal{H}$, the rate 
\begin{equation}
\label{equ:bar_R}
\bar{R}=R\left( \mathbf{w}^{\star},\mathbf{h}^{\star} \right)
\end{equation}
satisfies the constraint \eqref{equ:outage_constraint}. In fact, since $\left( \mathbf{w}^{\star},\mathbf{h}^{\star} \right)$ is a solution of problem \eqref{equ:problem_robustbf}, we have the inequality
\begin{equation}
	R( \mathbf{w}^{\star}, \mathbf{h}) \ge \bar{R},\mathrm{~for~all}~\mathbf{h}\in \mathcal{H}, 
\end{equation}
which implies the inequality
\begin{equation}
	\mathrm{Pr}\left[ R\left( \mathbf{w}^{\star},{\mathbf{h}} \right) \ge \bar{R} \right] \ge \mathrm{Pr}\left[ \mathbf{h}\in \mathcal{H} \right] \ge 1-\alpha. 
\end{equation}

Problem \eqref{equ:problem_robustbf} is tackled as detailed in Section \ref{sec:robustBF}.

\vspace{-5pt}
\section{Conformal Channel Estimation}
\label{sec:CE_CP}
In this section, we introduce a scheme that leverages CP to obtain a set $\mathcal{H}$ of estimated channel that ensures the coverage constraint \eqref{equ:not_in_set}.

\subsection{Calibration Data}
To produce the uncertainty set $\mathcal{H}$, we assume that the BS has access to a data set $\mathcal{D} =\left\{ \left( \mathbf{y}_i,\mathbf{h}_i \right) \right\} _{i\in \mathcal{I}}$, where each data point $\left( \mathbf{y}_i,\mathbf{h}_i \right)$ represents received signal $\mathbf{y}_i$ in \eqref{equ:simo_ce} and the corresponding CSI $\mathbf{h}_i$. The data points in the set $\mathcal{D}$ are assumed to follow the same unknown distribution underlying the generation of the channel $\mathbf{h}$ and received signal $\mathbf{y}$ in \eqref{equ:simo_ce}.
The dataset $\mathcal{D} $ is partitioned into a training set $\mathcal{D} ^{\mathrm{tr}}=\{\left( \mathbf{y}_i,\mathbf{h}_i \right)\} _{i\in \mathcal{I} ^{\mathrm{tr}}}$ and a calibration dataset $\mathcal{D} ^{\mathrm{cal}}=\{\left( \mathbf{y}_i,\mathbf{h}_i \right)\} _{i\in \mathcal{I} ^{\mathrm{cal}}}$, so that $\mathcal{I} ^{\mathrm{tr}}\cup \mathcal{I} ^{\mathrm{cal}}=\mathcal{I}$ and $\mathcal{I} ^{\mathrm{tr}}\cap \mathcal{I} ^{\mathrm{cal}}=\emptyset$.
As described below, the training set $\mathcal{D} ^{\mathrm{tr}}$ is used to train a channel estimator for the CSI $\hat{\mathbf{h}}$ given the received signal $\mathbf{y}$, while the calibration set $\mathcal{D} ^{\mathrm{cal}}$ is used to evaluate the prediction set $\mathcal{H}$.

\subsection{VAE-Based Generative Channel Model}
\label{subsec:VAE}
\begin{figure}[!t]
	\centering
	\includegraphics[width=0.7\columnwidth]{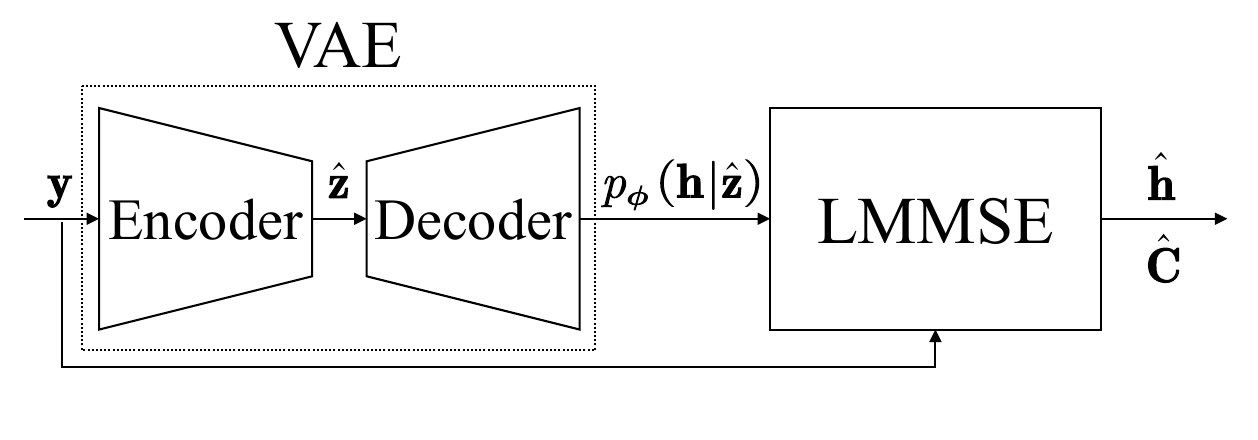}
	\caption{VAE-based channel estimator \cite{baur2024leveraging}.}
	\label{fig:VAE}
	\vspace{-15pt}
\end{figure}

Following \cite{baur2024leveraging}, we leverage a variational autoencoder (VAE)-based channel estimator. As depicted in Fig. \ref{fig:VAE}, the VAE-based estimator takes the signal $\mathbf{y}$ in \eqref{equ:simo_ce} as input, and it outputs an estimate of the channel distribution. The VAE consists of an encoder $q_{\boldsymbol{\psi }}(\mathbf{z}|\mathbf{y})=\mathcal{N} \left( \boldsymbol{\mu }_{\boldsymbol{\psi }}( \mathbf{y} ) ,\mathrm{Diag}\left( \boldsymbol{c}_{\boldsymbol{\psi }}( \mathbf{y} ) \right) \right) $,
mapping vector $\mathbf{y} $ to a latent vector $\mathbf{z}\in\mathbb{R}^{L}$, and a decoder 
$p_{\boldsymbol{\phi }}(\mathbf{h}|\mathbf{z})= \mathcal{CN} \left( \boldsymbol{\mu }_{\boldsymbol{\phi }}( \mathbf{z} ) ,\mathbf{C}_{\boldsymbol{\phi }}( \mathbf{z}) \right)$,
mapping latent vector $\mathbf{z}$ to channel $\mathbf{h}$, where ${\boldsymbol{\psi }}$ and ${\boldsymbol{\phi }}$ are trainable parameters for the encoder and decoder neural networks, respectively. Equivalently, the encoder maps the input data $\mathbf{y}$ into the latent variable as 
\begin{equation}
\mathbf{z}=\boldsymbol{\mu }_{\boldsymbol{\psi }}\left( \mathbf{y} \right) +\boldsymbol{c}_{\boldsymbol{\psi }}\left( \mathbf{y} \right) \odot \boldsymbol{\delta },
\end{equation}
where $\boldsymbol{\delta } \!\!\sim \!\!\mathcal{N} (\mathbf{0},\mathbf{I})$. The VAE parameters ${\boldsymbol{\psi }}$ and ${\boldsymbol{\phi }}$ are trained using standard methods based on a data set of channel realizations. 

Given a trained VAE, for a received signal $\mathbf{y}$ in \eqref{equ:simo_ce}, the posterior distribution $p(\mathbf{h}|\mathbf{y})$ can be approximated as follows \cite{baur2024leveraging}. First, we compute the maximum a posteriori (MAP) estimate $\hat{\mathbf{z}}=\boldsymbol{\mu }_{\boldsymbol{\psi }}(\mathbf{y})
$ for the latent vector under the encoder model. Then, the channel posterior distribution is estimated by using the decoding distribution $p_{\boldsymbol{\phi }}(\mathbf{h}|\hat{\mathbf{z}})$ as prior, yielding the approximation 
$p_{\boldsymbol{\phi }}(\mathbf{h}|\mathbf{y})\approx p_{\boldsymbol{\phi }}(\mathbf{h}|\hat{\mathbf{z}},\mathbf{y})=\mathcal{C} \mathcal{N} ( \hat{\mathbf{h}},\hat{\mathbf{C}} )$,
where 
\begin{subequations}
	\label{equ:lmmse_distribution}
	\begin{equation}
		\label{equ:lmmse_mean}
		\hat{\mathbf{h}}=\mathbf{y}-\sigma ^2( \mathbf{C}_{\boldsymbol{\phi }}( \mathbf{z} ) +\sigma ^2\mathbf{I}) ^{\!-1}( \mathbf{y}-\boldsymbol{\mu }_{\boldsymbol{\phi }}( \mathbf{z} )), 
	\end{equation}
	\begin{equation}
		\label{equ:lmmse_var}
		\hat{\mathbf{C}} =\sigma ^2\left( \mathbf{C}_{\boldsymbol{\phi }}\left( \mathbf{z} \right) +\sigma ^2\mathbf{I} \right) ^{-1}\mathbf{C}_{\boldsymbol{\phi }}\left( \mathbf{z} \right).
	\end{equation}
\end{subequations}
correspond to the linear minimum mean square error (LMMSE) estimate of the channel based on prior $p_{\boldsymbol{\phi }}(\mathbf{h}|\hat{\mathbf{z}})$ and received signal \eqref{equ:simo_ce}, and to the corresponding error covariance matrix, respectively.

Since this is an approximate posterior distribution, the set
\begin{equation}
\label{equ:naive_set} 
\mathop{\rm{argmin}}\limits_{\mathcal{H}\subseteq \mathbb{C}^{N \times 1}}\ |\mathcal{H}| \nonumber
\text{ s.t. } \int_{\mathbf{h} \in \mathcal{H}} p_{\boldsymbol{\phi}}(\mathbf{h}|\mathbf{y}) \mathrm{d}\mathbf{h} \geq 1-\alpha,
\end{equation}
which covers $1-\alpha$ of the mass of the approximate posterior, generally does not satisfy the coverage condition \eqref{equ:not_in_set}. This issue is addressed next via CP.
\subsection{Calibration via Conformal Prediction}
\label{subsec:CP}
In this subsection, we explain how to apply CP to calibrate the uncertainty set in \eqref{equ:naive_set}.
Given the estimated \eqref{equ:lmmse_distribution}, we first compute a score for each $i$-th calibration data pair $\left( \mathbf{y}_i,\mathbf{h}_i \right) $ as 
\begin{equation}
\label{equ:score}
s( \mathbf{y}_i,\mathbf{h}_i) =\| \mathbf{h}_i-\hat{\mathbf{h}}_i \| , \mathrm{~for~all}~i\in \mathcal{I}^{\mathrm{cal}}.
\end{equation}
The function $s(\mathbf{y}_i,\mathbf{h}_i)$ in \eqref{equ:score} measures the residual error between the true CSI $\mathbf{h}_i$ and the estimated CSI $\hat{\mathbf{h}}_i$ in \eqref{equ:lmmse_mean}. Subsequently, the prediction set is constructed as
\begin{equation}
\label{equ:cp_set}
\mathcal{H} ^{\mathrm{CP}}(\mathbf{y})=\left\{ \mathbf{h}\in \mathbb{C} ^{N \times 1}:s\left( \mathbf{y},\mathbf{h} \right) \le q \right\} ,
\end{equation}
thus including all channel vectors whose score is smaller than a threshold ${q}$. 

The threshold ${q}$ is finally evaluated as the $\left( \left| \mathcal{I} ^{\mathrm{cal}} \right|\!+\!1 \right)\! \left( 1\!-\!\alpha \right) \!/\!\left| \mathcal{I} ^{\mathrm{cal}}\right|$-quantile, i.e., the $\lceil \!\left( \left| \mathcal{I} ^{\mathrm{cal}} \right|\!+\!1 \right)\! \left( 1\!-\!\alpha \right)\! \rceil /\left| \mathcal{I} ^{\mathrm{cal}} \right| $-th smallest value, of the calibration scores $\left\{ s_i \right\} _{i\in \mathcal{I} ^{\mathrm{cal}}}$ \cite{angelopoulos2021gentle}.
Thanks to the properties of CP \cite{angelopoulos2021gentle}, assuming that the channel realizations in the calibration dataset $\mathcal{D}^{\mathrm{cal}}$ and the current channel $\mathbf{h}$ are drawn i.i.d. from the same (unknown) distribution, the prediction set in \eqref{equ:cp_set} guarantees the desired coverage of $1-\alpha$.

\section{Min-Max Robust Optimization}
\label{sec:robustBF}

In this section, we discuss how to address the min-max problem \eqref{equ:problem_robustbf} given the uncertainty set $\mathcal{H}^{\mathrm{CP}}$ in \eqref{equ:cp_set} following the approach in \cite{xiang2012robust}.

To start, we tackle the inner minimization in \eqref{equ:obj_robustbf}. It is observed that in the low-SNR regime in which the set $\mathcal{H}^{\mathrm{CP}}$ in \eqref{equ:cp_set} includes the channel $\mathbf{h}=\mathbf{0}$, i.e., if $\| \hat{\mathbf{h}} \| \le q$, the optimal value of the inner optimization in \eqref{equ:obj_robustbf} reduces to zero for any beamforming vector $\mathbf{w}$. In this case, the transmission rate \eqref{equ:bar_R}, $R\left( \mathbf{w}^{\star},\mathbf{h}^{\star}=\mathbf{0} \right) 
$, equals zero.
Otherwise, given the condition $\| \hat{\mathbf{h}} \| > q$, the inner minimization of problem \eqref{equ:problem_robustbf} can be solved as in \cite{xiang2012robust}, yielding
\begin{align}
\label{equ:inner_min}
\min_{\mathbf{h}\in \mathcal{H} ^{\mathrm{CP}}} \left| \mathbf{h}^{\mathsf{H}}\mathbf{w} \right|^2=\begin{cases}
	0&,~\text{if}~\| \hat{\mathbf{h}} \| \le q\\
	\left( | \hat{\mathbf{h}}^{\mathsf{H}}\mathbf{w} |-q\left\| \mathbf{w} \right\| \right) ^2&,~\text{if}~\| \hat{\mathbf{h}} \| >q\\
\end{cases}.
\end{align}
Furthermore, for the case $\|\hat{\mathbf{h}} \| >q$, the optimized vector $\mathbf{h}^{\star}$ is expressed as
\begin{equation}
\mathbf{h}^{\star}=\hat{\mathbf{h}}-\frac{\mathbf{w}^{\star}}{\left\| \mathbf{w}^{\star} \right\|}qe^{-j\theta},
\end{equation}
where $\theta =\mathrm{arccos}(| \hat{\mathbf{h}}^{\mathsf{H}}\mathbf{w}^{\star} |/\| \hat{\mathbf{h}} \| \| \mathbf{w}^{\star} \| 
) 
$.

Based on the result in \eqref{equ:inner_min}, assuming $\|\hat{\mathbf{h}} \| >q$, the optimal beamforming vector $\mathbf{w}^{\star}$ is finally obtained as 
\begin{equation}
\label{equ:w_opt}
\mathbf{w}^{\star}=\sqrt{P}\frac{\hat{\mathbf{h}}}{\| \hat{\mathbf{h}} \|}.
\end{equation}

\section{Numerical Results}
\label{sec:numerical_results}
In this section, we present numerical results to demonstrate the benefits and effectiveness of the proposed conformal robust beamforming framework.
\subsection{Simulation Setup}
We investigate a MISO system with $N=32$ antennas at the BS. 
Following the 3rd Generation Partnership Project (3GPP) standard \cite{3gpp_tr_25_996_v16}, the channel is simulated as the multivariate Gaussian vector $\mathbf{h}\sim\mathcal{CN} \left( \mathbf{0},\mathbf{C}_h\right)$,
where the channel covariance $\mathbf{C}_h$ is expressed as 
\begin{equation}
	\mathbf{C}_h=\int_{-\pi}^{\pi}{g\left( \varphi\right)}\mathbf{a}\left( \varphi \right) \mathbf{a}\left( \varphi \right)^{\mathsf{H}} d\varphi, 
\end{equation}
with $g\left( \varphi \right) \in \mathbb{R}^{+}$ representing the power angular spectrum and $\mathbf{a}\left( \varphi \right) \in\mathbb{C}^{N\times 1}$ being the array response vector for angle $\varphi $. The power angular spectrum $g\left( \varphi \right)$ is characterized following the 3GPP urban macrocell spatial channel model. Furthermore, the array response $\mathbf{a}\left( \varphi \right)$ is that of a linear array with a half-wavelength antenna spacing.

All channel samples are generated following the 3GPP model. The channel power is $\mathbb{E} [ \left\| \mathbf{h} \right\| ^2 ] =N$. Accordingly, the SNR in the channel training stage is defined as $\mathrm{SNR}^{\mathrm{tr}}=N/\gamma ^2$, while the SNR of data transmission is $\mathrm{SNR}=NP/\sigma ^2$.
The VAE model is implemented following \cite{baur2024leveraging} with latent dimension $L=32$. A total of $\left|\mathcal{I}^{\mathrm{tr}}\right|= 180000$ training samples for a range of $\mathrm{SNR}^{\mathrm{tr}}$ from $-5$ to $45$ dB are used for channel training, while $\left|\mathcal{I}^{\mathrm{cal}}\right|= 100$ data pairs
are used for calibration.

As a baseline, we consider the conventional method that directly adopts the ($1-\alpha$) uncertainty as in \eqref{equ:naive_set} to address the min-max problem \eqref{equ:problem_robustbf}. 

\subsection{Performance Metrics}
For all the considered schemes, we report the empirical channel coverage, the empirical outage probability, and the empirical average rate evaluated on the test dataset $\mathcal{D} ^{\mathrm{te}}=\left\{ \left( \mathbf{y}_i,\mathbf{h}_i \right) \right\} _{i\in \mathcal{I} ^{\mathrm{te}}}$ of size $\left|\mathcal{I} ^{\mathrm{te}}\right|=100$. 
\subsubsection*{\normalfont\arabic{subsubsection}) \itshape Empirical Channel Coverage}
\addtocounter{subsubsection}{1} 
The empirical coverage measures the fraction of times in which the true channel is included in the prediction set $\mathcal{H}(\mathbf{y})$, i.e.,
\begin{equation}
\label{equ:metric_coverage}
\mathrm{Channel~Coverage}=\frac{1}{|\mathcal{I} ^{\mathrm{te}}|}\sum_{i\in \mathcal{I} ^{\mathrm{te}}}{\mathbbm{1}}(\mathbf{h}_i\in \mathcal{H} (\mathbf{y}_i)),
\end{equation} 
where $\mathbbm{1}$ denotes the indicator function ($\mathbbm{1} \left( \mathrm{true} \right) =1
$ and $\mathbbm{1} \left( \mathrm{false} \right) =0
$). 
\subsubsection*{\normalfont\arabic{subsubsection}) \itshape Empirical Outage Probability}
\addtocounter{subsubsection}{1} 
The empirical outage probability quantifies the fraction of times that the achievable rate on true channel realization $\mathbf{h}_i$ falls below the target rate $\bar{R}_i$ in \eqref{equ:bar_R}, i.e.
\begin{equation}
	\label{equ:metric_outage}
	\mathrm{Outage~Probability}=\frac{1}{|\mathcal{I} ^{\mathrm{te}}|}\sum_{i\in \mathcal{I} ^{\mathrm{te}}}{{\mathbbm{1}}(R\left( \mathbf{w}_{i}^{\star},\mathbf{h}_i \right) <\bar{R }_i)}.
\end{equation}

\subsubsection*{\normalfont\arabic{subsubsection}) \itshape Empirical Average Rate}
\addtocounter{subsubsection}{1} 
The empirical average rate is evaluated by averaging the transmission rate in \eqref{equ:bar_R}, i.e.,
\begin{equation}
\label{equ:metric_rate}
\mathrm{Average~Rate}=\frac{1}{|\mathcal{I} ^{\mathrm{te}}|}\sum_{i\in \mathcal{I} ^{\mathrm{te}}}{\bar{R }_i},
\end{equation}

We average all metrics over 200 experiments, each corresponding to independent draws of the calibration and test data set pair $\left\{ \mathcal{D} ^{\mathrm{cal}},\mathcal{D} ^{\mathrm{te}} \right\} $. 

\subsection{Performance Analysis}
%

\begin{figure*}[t]
	\centering
	\begin{subfigure}{0.32\textwidth}
		\centering
		\includegraphics[width=\textwidth]{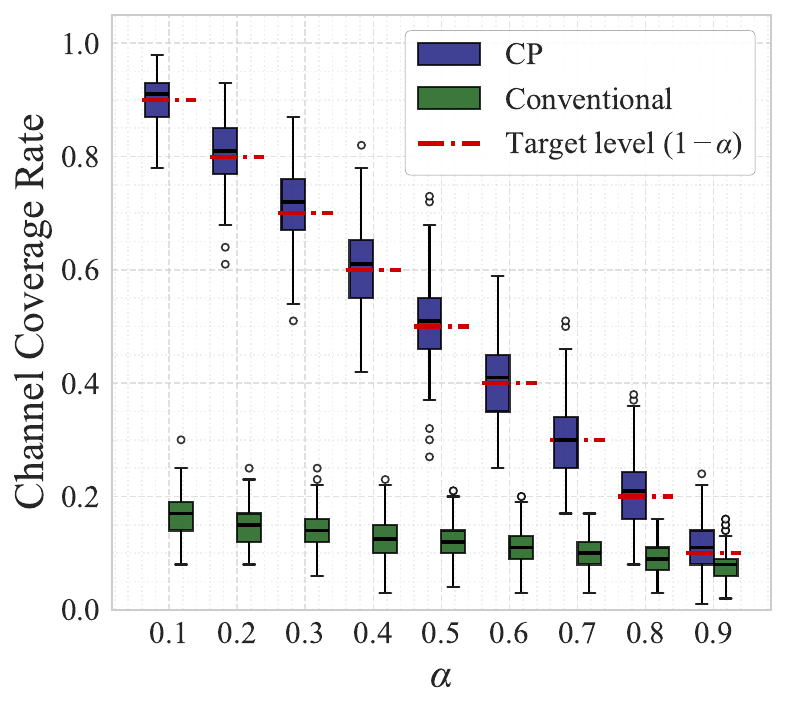}
		\vspace{-16pt}
		\caption{Channel coverage}
		\label{subfig:coverage_snr-5}
	\end{subfigure}
	\hfill
	\begin{subfigure}{0.32\textwidth}
		\centering
		\includegraphics[width=\textwidth]{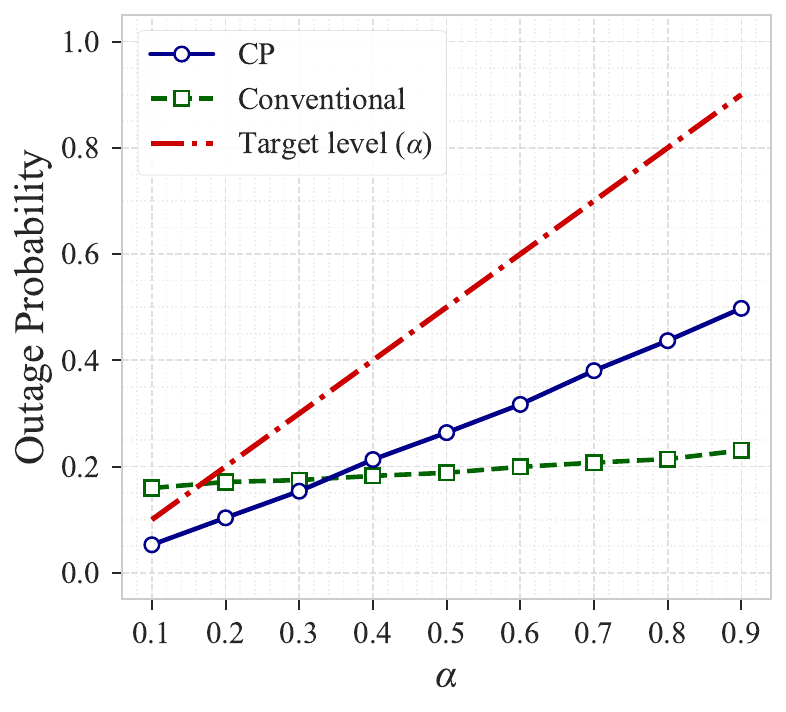}
		\vspace{-16pt}
		\caption{Outage probability}
		\label{subfig:outage_snr-5}
	\end{subfigure}
	\hfill
	\begin{subfigure}{0.32\textwidth}
		\centering
		\includegraphics[width=\textwidth]{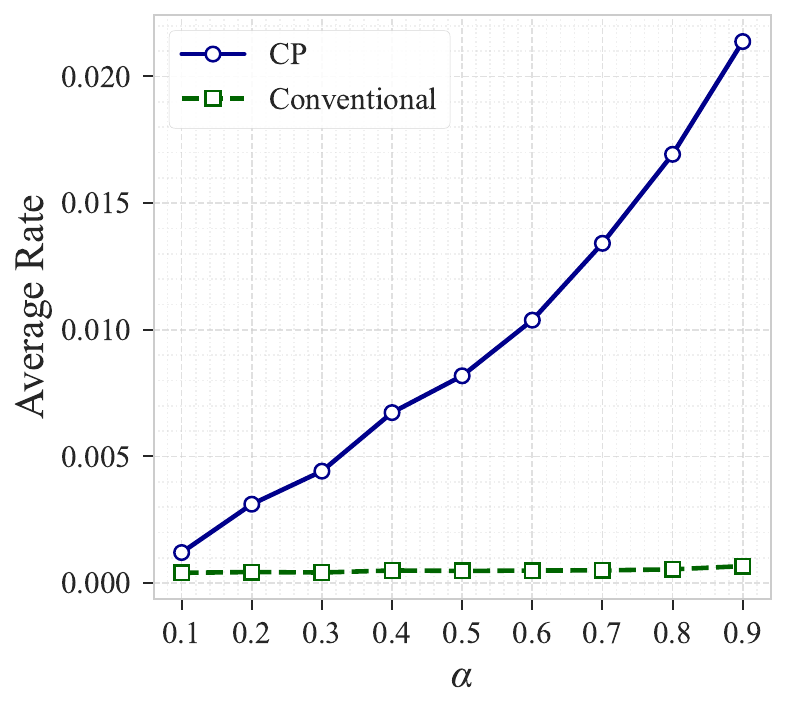}
		\vspace{-16pt}
		\caption{Average rate}
		\label{subfig:rate_snr-5}
	\end{subfigure}
	\caption{Channel coverage \eqref{equ:metric_coverage}, outage probability \eqref{equ:metric_outage} and average rate \eqref{equ:metric_rate} as a function of the target outage probability $\alpha$ in \eqref{equ:outage_constraint}
		with $\mathrm{SNR}^{\mathrm{tr}}=-5$ dB and $\mathrm{SNR}=-5$ dB. The results are averaged over 200 experiments.}
	\label{fig:snr-20}
\end{figure*}

Fig. \ref{fig:snr-20} compares the performance of the proposed CP-based approach and of the mentioned baseline for $\mathrm{SNR}^{\mathrm{tr}}=\mathrm{SNR}= -5$ dB. In this low-SNR scenario, as shown in Fig. \ref{subfig:coverage_snr-5}, the conventional method consistently fails to satisfy the coverage requirements across different values of the target miscoverage probability $\alpha$, while the proposed CP method reliably maintains the target channel coverage rate. As reported in Fig. \ref{subfig:outage_snr-5}, 
the inadequate channel coverage of conventional method results in violations of the outage probability constraint in settings with strict target outage probabilities $\alpha$. In contrast, the proposed CP-based scheme can always ensure the outage constraint. 

The average rate performance in Fig. \ref{subfig:rate_snr-5} further demonstrates the effectiveness of the proposed CP-based scheme on robust beamforming. In fact, the conventional method yields a near-zero rate due to the exceedingly large uncertainty sets \eqref{equ:naive_set}, which cannot adapt to the outage probability $\alpha$. In contrast, the proposed CP-based scheme is able to adaptively calibrate the uncertainty set, obtaining the increasing average rates as the target outage probability $\alpha$ increases.
\begin{figure*}[t]
	\centering
	\vspace{-6pt}
	\begin{subfigure}{0.32\textwidth}
		\centering
		\includegraphics[width=\textwidth]{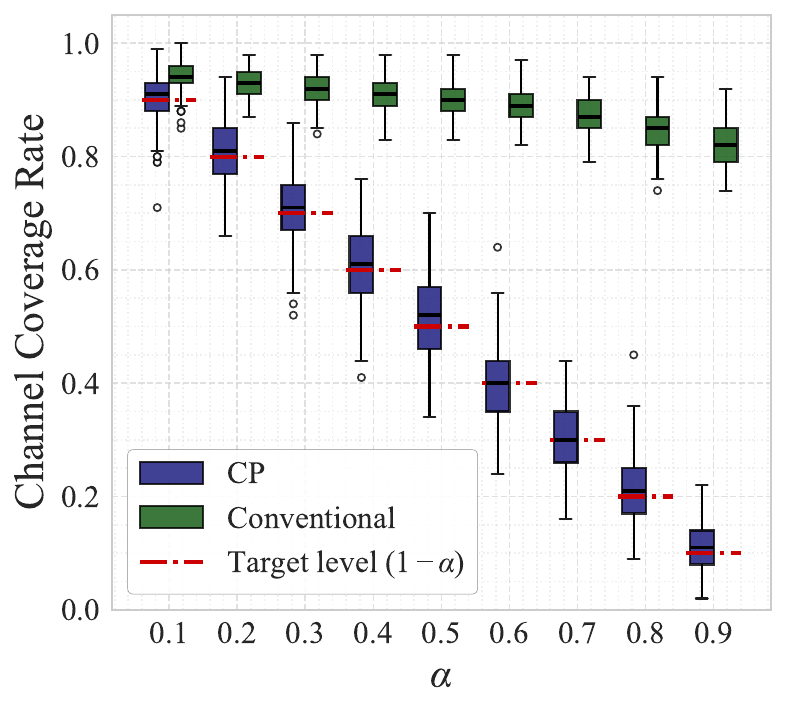}
		\vspace{-16pt} 
		\caption{Channel coverage}
		\label{subfig:coverage_snr25}
	\end{subfigure}
	\begin{subfigure}{0.32\textwidth}
		\centering
		\includegraphics[width=\textwidth]{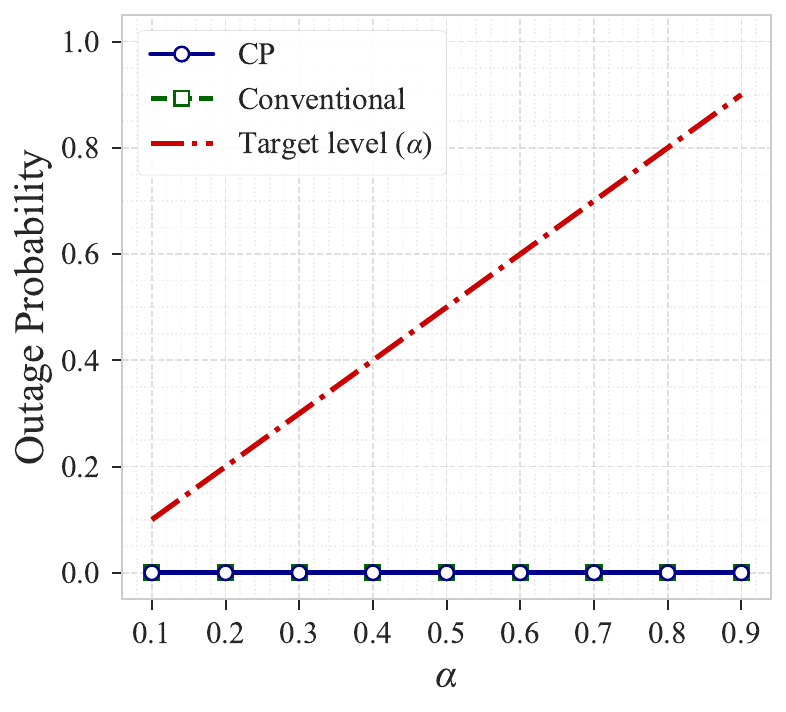}
		\vspace{-16pt} 
		\caption{Outage probability}
		\label{subfig:outage_snr25}
	\end{subfigure}
	\hfill
	\begin{subfigure}{0.32\textwidth}
		\centering
		\includegraphics[width=\textwidth]{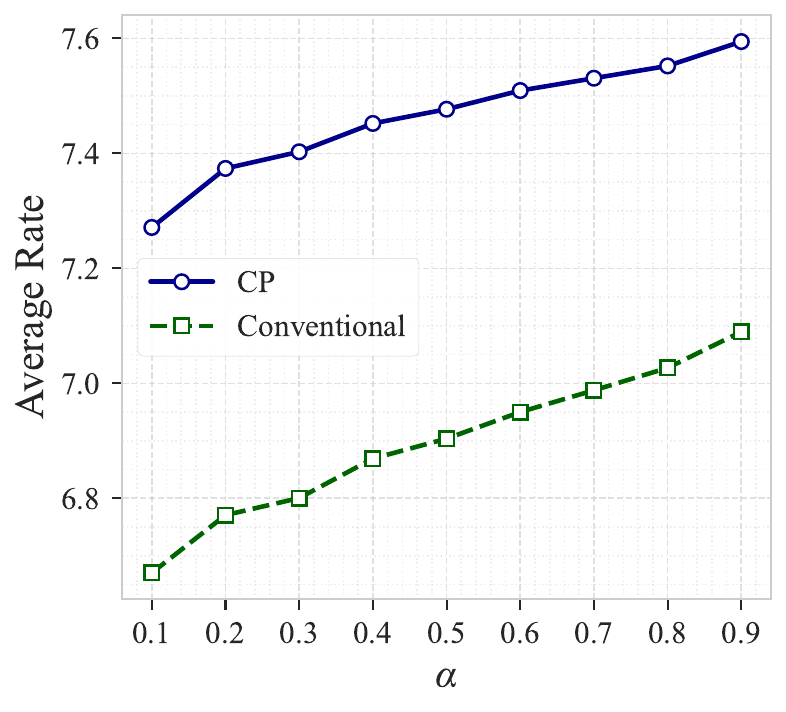}
		\vspace{-16pt} 
		\caption{Average rate}
		\label{subfig:rate_snr25}
	\end{subfigure}
	\hfill
	\caption{Channel coverage \eqref{equ:metric_coverage}, outage probability \eqref{equ:metric_outage} and average rate \eqref{equ:metric_rate} as a function of the target outage probability $\alpha$ in \eqref{equ:outage_constraint}
		with $\mathrm{SNR}^{\mathrm{tr}}=25$ dB and $\mathrm{SNR}=25$ dB. The results are averaged over 200 experiments.}
	\label{fig:snr25}
	\vspace{-14pt} 
\end{figure*}

As illustrated in Fig. \ref{subfig:coverage_snr25}, in high-SNR conditions, here $\mathrm{SNR}^{\mathrm{tr}}= 25$ dB, both the conventional method and the proposed CP-based scheme satisfy the coverage guarantee requirement, attaining near-zero outage probabilities as seen in Fig. \ref{subfig:outage_snr25}. However, the conventional method adopts an overly conservative set. Accordingly, as depicted in Fig. \ref{subfig:rate_snr25}, the proposed CP-based approach consistently achieves superior average rates compared to the conventional method regardless of the value of $\alpha$.

\vspace{-5pt}
\section{Conclusion}
\label{sec:conclusion}
This paper has proposed a novel two-step conformal robust beamforming framework for maximizing the system's achievable rate under outage probability constraints that hold irrespective of making any assumptions about the underlying channel distribution. The approach begins by constructing a reliable channel uncertainty set using conformal prediction, which provides finite-sample coverage guarantees. In the second step, the framework addresses a min-max beamforming problem over the learned uncertainty set, enabling robust performance even under CSI uncertainty. Numerical results have validated the effectiveness and computational efficiency of the proposed method across a range of SNR conditions, demonstrating its potential for practical deployment in next-generation wireless systems. Future work may address multi-user and multi-carrier settings.
 
%
	
\bibliographystyle{IEEEtran}
\bibliography{./bibtex/IEEEabrv,./bibtex/IEEEreference}

\end{document}